\documentclass[12pt]{amsart}
\usepackage{graphicx}
\usepackage{geometry}
\usepackage{amssymb}
\usepackage{setspace}
\usepackage[dvipsnames]{pstricks}
\usepackage{pst-node}
\usepackage{pst-slpe}
\usepackage{pst-tree}
\usepackage{appendix}
\usepackage{multirow}
\usepackage{ulem}
\usepackage{setspace}
\usepackage{natbib}
\usepackage{pstricks}
\usepackage{enumitem}
\usepackage{fancybox}
\usepackage{comment}
\usepackage{epstopdf}
\usepackage{bigints}
\usepackage{tikz}
\usetikzlibrary{shapes,snakes}
\usetikzlibrary{calc,intersections,through,backgrounds,patterns}
\usepackage{pgfplots}
\pgfplotsset{compat=newest}
\usepackage{float}
\usepackage{amsfonts}
\usepackage{amsmath}
\usepackage{appendix}
\usepackage{multirow}
\usepackage{ulem}
\usepackage{setspace}
\usepackage{natbib}
\usepackage{pstricks}
\usepackage{pst-node}
\usepackage{enumitem}
\usepackage{fancybox}
\usepackage{comment}
\usepackage{bigints}
\everymath{\displaystyle}

\usepackage{tikz}
\usetikzlibrary{shapes,snakes}
\usetikzlibrary{calc,intersections,through,backgrounds,patterns}
\pgfplotsset{compat=newest}

\newtheorem{proposition}{Proposition}

\usepackage{caption}
\usepackage{subcaption}
\usepackage{tikz}
\usetikzlibrary{decorations.pathreplacing} 
\newcommand{\argmax}{\operatornamewithlimits{argmax}}
\begin{document}

\title[]{Contract Design with Costly Convex Self-Control$^\ast$}
\author[]{Yusufcan Masatlioglu$^\S$}
\author[]{Daisuke Nakajima$^\dag$}
\author[]{Emre Ozdenoren$^\ddag$}
\thanks{$^\S$ University of Maryland, 3147E Tydings Hall, 7343 Preinkert Dr.,  College Park, MD 20742. E-mail: \texttt{yusufcan@umd.edu}}
\thanks{$^\dag$ Otaru University of Commerce. E-mail: 
\texttt{nakajima@res.otaru-uc.ac.jp.}}
\thanks{$^\ddag$ London Business School. Email:\texttt{eozdenoren@london.edu}}
\date{July, 2019}
\maketitle

In this note, we solve the pricing problem of a profit-maximizing monopolist who faces consumers with convex self-control preferences. The idea that self-control costs can be convex has been introduced by \cite{Fudenberg_Levine06}. Building on \cite{GP01}, \citet{NoorTakeoka:2010, NoorTakeoka:2015} provide axiomatic characterizations of convex self-control.  Monopolistic contracting with consumers who have self-control problems was studied by \cite{DellavignaMalmendier04}, \cite{Eliaz_Spiegler06}, \cite{Heidhues_Koszegi10}, among others. The previous literature has highlighted that the monopolist can exploit naive consumers by offering them indulging contracts with two alternatives. Our earlier work, \cite{MNOzdenoren19} builds on \cite{Eliaz_Spiegler06} and shows that when consumers have limited willpower preferences, monopolist offers compromising contracts with three alternatives. This note extends the solution of that problem to naive consumers with convex self-control preferences.

\section{Model}\label{sec:contractconvex}

We denote the finite set of alternatives available to
the monopolist by $X$. A contract $C$ is a menu of offers, where each offer is an alternative with an associated price, i.e., $C=\{(s,p(s)):s \in S \subset X\}$. We consider a two-period model of contracting between a
monopolist and a consumer. In the first period, the monopolist offers the consumer a contract $C$. The consumer can accept or reject the contract.
If the consumer accepts the contract, in the second period, he chooses an
offer from the contract and pays its price to the monopolist. If the
consumer rejects the contract, then he receives his outside option normalized
to zero. We assume that both parties are committed to the contract once
accepted. 

The monopolist's profit from selling alternative $s$ at price $p(s)$ is $p(s)-c\left(
s\right).$
The production cost is incurred only for the service that the
consumer chooses from the menu.

We assume that the consumer has costly convex self-control preferences where the cost function is given by $\varphi: \mathbb{R}^+ \rightarrow \mathbb{R}^+$. We assume that $\varphi$ is a convex, continuous, strictly increasing function such that $\varphi(0)=0$. We
assume that the consumer is naive in the sense that he believes he has no
self-control problem, i.e., he believes that from a contract $C$ he will choose the offer
$(s,p(s))$ that maximizes $U\left( s,p(s) \right) = u(s) - p(s) $.
In reality, the consumer's second period choices are governed by the costly convex self-control model. This means that from $C$ the consumer chooses the offer
$(s,p(s))$ that maximizes $U\left( s,p(s) \right)-\varphi\left(\max_{(s',p(s')) \in C}V\left( s',p(s')\right)-V\left( s,p(s) \right) \right)$ where $V(s,p(s))=v(s)-p(s)$.

To simplify the analysis, we assume that $u-c$ and $v-c$ have unique maximizers $x^u$  and $x^v$ in $A$. In other words, $x^u$  and $x^v$ are the most efficient alternatives with respect to $u$ and $v$. To make the problem interesting, we assume that $x^u \neq x^v$. We define the difference between the temptation value and the
utility value of an alternative $s$ as its \textit{excess temptation} and
denote it by $e\left( s\right) \equiv
v\left( s\right) -u\left( s\right) .$ We further assume that $e$ has a unique maximizer, $z^*$, and a unique minimizer, $y^*$, in $A$.  

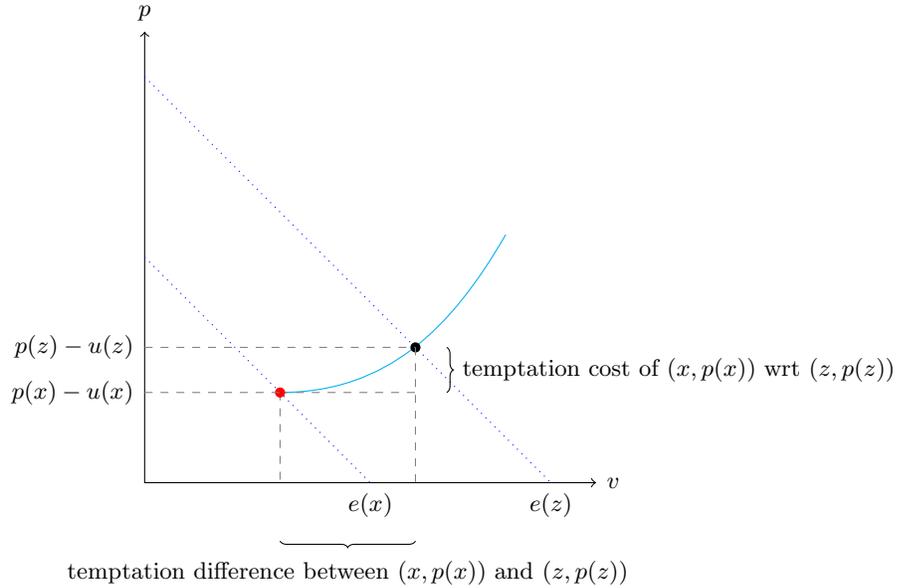
\begin{figure}[h!]  
\begin{center}
\begin{tikzpicture}[scale=1.2] 
\draw [<->] (0,5) -- (0,0) -- (5,0);
\node [right] at (5,0) {\scriptsize{$v$}};
\node [above] at (0,5) {\scriptsize{$p$}};
\draw [blue, dotted] (0,2.5) to (2.5,0);
\draw [blue, dotted] (0,4.5) to (4.5,0);
\node [below] at (4.5,0) {\scriptsize{$e(z)$}};
\node [below] at (2.5,0) {\scriptsize{$e(x)$}};






\draw[Cyan] (1.5,1) to [out=0,in=240] (4,2.75);

\draw[fill] (3,1.5) circle [radius=0.05];
\draw[dashed, gray] (3,1.5) -- (3,0);
\draw[dashed, gray] (1.5,1) -- (3,1);

\draw[dashed, gray] (1.5,1) -- (1.5,0);


\draw[fill=red, red] (1.5,1) circle [radius=0.05];

\draw[decorate,decoration={brace,raise=5pt}] (3.2,1.5) -- (3.2,1);
\node [right] at (3.4,1.25) {\scriptsize{temptation cost of $(x,p(x))$ wrt $(z,p(z))$ }};

\draw[decorate,decoration={brace, mirror, raise=5pt}] (1.5,-0.5) -- (3,-.5);
\node [below] at (2.25,-.75) {\scriptsize{temptation difference between  $(x,p(x))$ and $(z,p(z))$ }};


\draw[dashed, gray] (0,1.5) -- (3,1.5) ;
\node [left] at (0,1.5)  {\scriptsize{$p(z)-u(z)$}};

\draw[dashed, gray] (0,1) -- (1.5,1) ;
\node [left] at (0,1)  {\scriptsize{$p(x)-u(x)$}};
\end{tikzpicture}
\caption{Alternatives indifferent to $(x,p(x))$ when they are the most tempting in a menu.} 
\label{fig:convex_setup}
\end{center}
\end{figure}

Next we introduce Figure \ref{fig:convex_setup} that we use extensively in the analysis below. Fix an arbitrary alternative $(x,p(x))$ with excess temptation $e(x)$ (the red point). For any $v \geq v(x)-p(x)$ the blue line corresponds to $p(x)-u(x)+\phi(v-v(x)+p(x))$.  Thus if $(x,p(x))$ is in a menu where the temptation value of the most tempting alternative is $v$, the light blue line gives the negative of the overall utility of $(x,p(x))$. Suppose $(x,p(x))$ is in a menu where the most tempting alternative is $(z,p(z))$ with excess temptation $e(z)$ (the black point). Then we have $v=v(z)-p(z)$ and the negative of the overall utility of $(x,p(x))$ is $p(x)-u(x)+\phi(v(z)-p(z)-v(x)+p(x))$. At the same time, from the iso-$e$ line we see that $p(z)-u(z)=p(x)-u(x)+\phi(v(z)-p(z)-v(x)+p(x))$, which is the negative of the overall utility of $(z,p(z))$ (which has no temptation cost). But this means that $(x,p(x))$ and $(z,p(z))$ are indifferent.   Hence, the blue line traces all the alternatives which are indifferent to $(x,p(x))$ when they are the most tempting alternative in the menu. 

\subsection{Solving for Optimal Contract}\label{sec:solveconvex}

First, we look for the revenue maximizing contract that sells an alternative $x$. W.l.o.g. we will restrict attention to three types of contracts: commitment, indulging and compromising.

\textsc{Commitment Contract:} In this case, the monopolist sells $x$ at price $p(x) = u(x)$.

\textsc{Indulging Contract:} Monopolist chooses $x$, $p(x)$, $y$ and $p(y)$
to maximize $p(x)$ subject to
\[
\begin{array}{c}
u\left(y\right)-p(y)\geq0\\
u\left(x\right)-p(x)\geq u\left(y\right)-p(y)-\varphi\left(v\left(x\right)-p(x)-v\left(y\right)+p(y)\right)
\end{array}
\]
For a given $y$, increasing $p(y)$ relaxes the second constraint, hence the first constraint must be binding. This implies that:
\[
u\left(x\right)-p(x)\geq -\varphi\left(v\left(x\right)-p(x)-e(y)\right).
\]
This means that we choose $y$ to minimize $e(y)$. Hence $y^*$ and $p(y^*)=u(y^*)$. The second constraint is also binding implying that:

\begin{equation}\label{C_ind}
p^{ind}(x)=u\left(x\right)+\varphi ( v (x)-p(x)-e(y^*)).
\end{equation}

\textsc{Compromising Contract:}
Monopolist chooses $p(x)$, $y$, $p(y)$,
$z$, $p(z)$ to maximize $p(x)$ subject to
{\small{\begin{eqnarray}
& u\left(y\right)-p(y)   \geq  0  \label{C-PC} \\
 & u\left(x\right)-p(x)-\varphi\left(v\left(z\right)-p(z)-v\left(x\right)+p(x)\right) \geq  u\left(y\right)-p(y)-\varphi\left(v\left(z\right)-p(z)-v\left(y\right)+p(y)\right) \label{C-xovery} \\
&u\left(x\right)-p(x)-\varphi\left(v\left(z\right)-p(z)-v\left(x\right)+p(x)\right) \geq  u\left(z\right)-p(z) \label{C-xoverz} 
\end{eqnarray}}}
Note that the consumer prefers $(x,p(x))$ to both $(y,p(y))$ and $(z,p(z))$. As usual $(y,p(y))$ is the bait. In the indulging contract $(x,p(x))$ has dual roles. It is both the chosen and the tempting alternative. In the compromising contract, the role of the tempting alternative is instead given to $(z,p(z))$. While we cannot solve the optimal compromising contract explicitly, we illustrate it graphically. This helps us to show that in this model the compromising contract  always  dominates the indulging contract for strictly convex cost functions.

\begin{figure}[h!]  
\begin{center}
\begin{tikzpicture}[scale=1.2] 
\draw [<->] (0,5) -- (0,0) -- (5,0);
\node [right] at (5,0) {\scriptsize{$v$}};
\node [above] at (0,5) {\scriptsize{$p$}};
\draw [blue, dotted] (0,.5) to (.5,0); 
\draw [blue, dotted] (0,2.5) to (2.5,0);
\draw [blue, dotted] (0,4.5) to (4.5,0);
\node [below] at (4.5,0) {\scriptsize{$e(z^*)$}};
\node [below] at (.5,0) {\scriptsize{$e(y^*)$}};
\node [below] at (2.5,0) {\scriptsize{$e(x)$}};


\draw[Cyan] (.5,0) to [out=0,in=240] (3,1.5);
\draw[Cyan] (3,1.5) to [out=60,in=250] (3.5,2.5);
\draw[Cyan] (1.5,1) to [out=0,in=240] (4,2.75);
\draw[fill] (0.5,0) circle [radius=0.05];
\draw[fill] (3,1.5) circle [radius=0.05];
\draw[dashed, gray] (3,1.5) -- (3,0);
\draw[dashed, gray] (1.5,1) -- (3,1);
\draw[fill=red, red] (2.06,.44) circle [radius=0.05];
\node [above] at (2.06,.44)  {\scriptsize{B}};
\draw[fill=red, red] (1.5,1) circle [radius=0.05];
\node [above right] at (1.5,1) {\scriptsize{C}};
\draw[fill=red, red] (2.5,0) circle [radius=0.05];
\node [above right] at (2.5,0) {\scriptsize{A}};
\draw[decorate,decoration={brace,raise=5pt}] (3.2,1.5) -- (3.2,1);
\node [right] at (3.4,1.25) {\scriptsize{temptation cost of $(x,p^{comp})$ wrt $(z^*,p(z^*))$ }};

\draw[decorate,decoration={brace,raise=5pt}] (3,1.5) -- (3,0);
\node [right] at (3.2,.75) {\scriptsize{temptation cost of $(y^*,p(y^*))$ wrt $(z^*,p(z^*))$ }};

\draw[dashed, gray] (0,1.5) -- (3,1.5) ;
\node [left] at (0,1.5)  {\scriptsize{$p(z^*)-u(z^*)$}};

\draw[dashed, gray] (0,1) -- (1.5,1) ;
\node [left] at (0,1)  {\scriptsize{$p^{comp}-u(x)$}};
\draw[dashed, gray] (0,.44) -- (2.06,.44);
\node [left] at (0,.44)  {\scriptsize{$p^{ind}-u(x)$}};
\end{tikzpicture}
\caption{Revenue maximizing commitment, indulging and compromising contracts for selling $x$.} 
\label{fig:convex}
\end{center}
\end{figure}
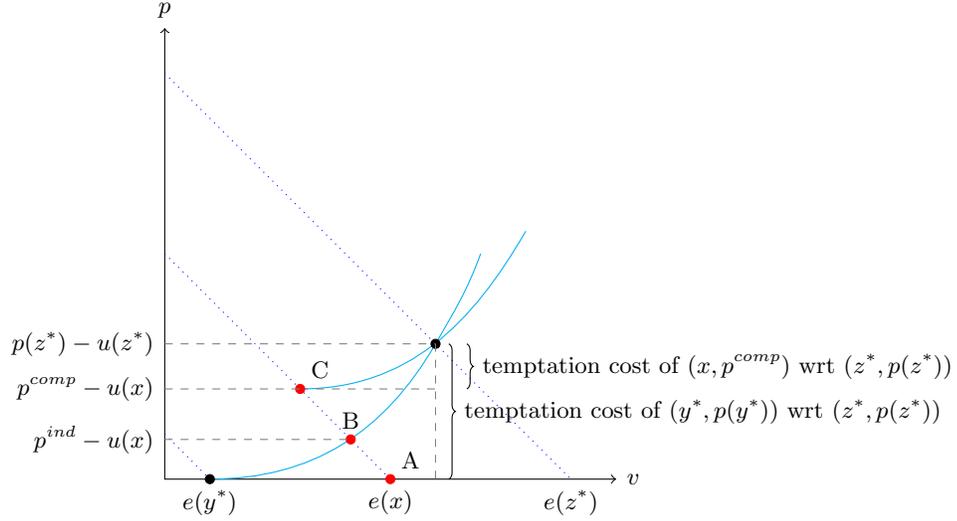

In this model, all three constraints are always binding and $y=y^*$ and $z=z^*$. Figure \ref{fig:convex} illustrates how to construct the optimal contract. First note that since constraint (\ref{C-PC}) is always binding and $u(y^*)=p(y^*)$. Since  (\ref{C-xovery}) and (\ref{C-xoverz}) bind, the consumer must be indifferent between all three contracts in the optimal contract.  This means that the monopolist (i) sets the price of $z^*$ as $p(z^*)$ such that $(y^*,u(y^*))$ and $(z^*,p(z^*))$ lie on the blue line starting from $(y^*,u(y^*))$, and  (ii) sets the price of $x$ as $p^{comp}$ such that  $(x,p^{comp})$ and $(z^*,p(z^*))$ lie on the blue line starting from  $(x,p^{comp})$.  Hence, the optimal contract is $\{(x,p^{comp}),(y^*,u(y^*)),(z^*,p(z^*))\}$ and the prices are given by two implicit equations:
\begin{equation}\label{eq:priceofz}
p(z^*)=u\left(z^*\right)+\varphi\left(v\left(z^*\right)-p(z^*)-e\left(y^*\right)\right)
\end{equation}
and \begin{equation}\label{eq:revmaxprice}
p^{comp}=u\left(x\right)+p(z^*)-u(z^*)-\varphi\left(v\left(z^*\right)-p(z^*)-v\left(x\right)+p^{comp}\right).
\end{equation}

To find the indulging contract to sell $x$, the monopoly needs to set the price of $x$ at $p^{ind}$ such that  $(x,p^{ind})$ and $(y^*,u(y^*))$ lie on the blue line starting from  $(y^*,p(y^*))$ (see Equation (\ref{C_ind})). As can be seen from the figure, the compromising contract generates  higher revenue  compared to both the indulging and the commitment contracts for each $x$. If the cost function is strictly convex, then it strictly dominates the others. 
 
Figure \ref{fig:optimal_comp} illustrates the optimal contract. Given the above discussion, we know that it is a compromising contract. We plot for each $x$ the optimal compromising contract selling $x$ which is the dark blue line in the figure. Since the cost function is convex, this line must be concave. The product sold in the optimal contract is illustrated by the point $D$, where the dark blue line is tangent to the iso-profit line. Recall that the iso-profit line has zero (infinite) slope at the point where it crosses the iso-$e$ line for $x^u$ ($x^v$). Since the slope of the blue line is strictly positive and finite, $D$ must have excess temptation strictly between $e(x^u)$ and $e(x^v)$.  Hence, the optimal contract always sacrifices efficiency for exploitation.

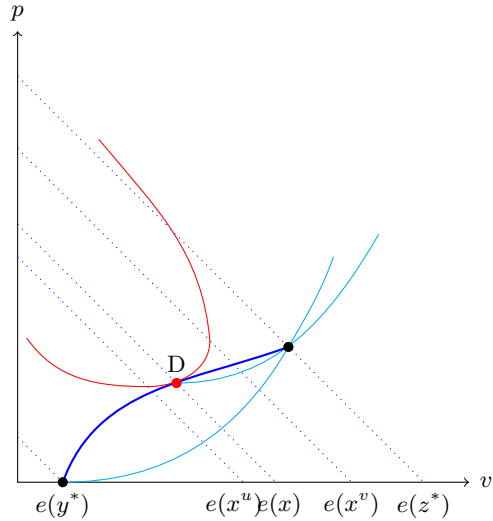
\begin{figure}[h!]  
\begin{center}
\begin{tikzpicture}[scale=1.2] 
\draw [<->] (0,5) -- (0,0) -- (5,0);
\node [right] at (5,0) {\scriptsize{$v$}};
\node [above] at (0,5) {\scriptsize{$p$}};
\draw [blue, dotted] (0,.5) to (.5,0); 
\draw [blue, dotted] (0,2.5) to (2.5,0);
\draw [blue, dotted] (0,2.86) to (2.86,0);
\draw [blue, dotted] (0,3.7) to (3.7,0);
\draw [blue, dotted] (0,4.5) to (4.5,0);
\node [below] at (4.5,0) {\scriptsize{$e(z^*)$}};
\node [below] at (.5,0) {\scriptsize{$e(y^*)$}};
\node [below] at (2.4,0) {\scriptsize{$e(x^u)$}};
\node [below] at (3.7,0) {\scriptsize{$e(x^v)$}};
\node [below] at (2.9,0) {\scriptsize{$e(x)$}};


\draw[Cyan] (.5,0) to [out=0,in=240] (3,1.5);
\draw[Cyan] (3,1.5) to [out=60,in=250] (3.5,2.5);
\draw[Cyan] (1.76,1.1) to [out=0,in=240] (4,2.75);
(1.76,1.1)

\draw[blue, thick] (.5,0) to [out=70,in=200] (3,1.5);


\draw[red] (1.44,1.06) to [out=0,in=-90] (2.13,1.57);
\draw[red] (2.13,1.57) to [out=95,in=310] (.9,3.8);
\draw[red] (1.44,1.06) to [out=180,in=305] (0.1,1.6);

\draw[fill] (0.5,0) circle [radius=0.05];
\draw[fill] (3,1.5) circle [radius=0.05];
\draw[fill=red, red] (1.76,1.1) circle [radius=0.05];
\node [above] at (1.76,1.1) {\scriptsize{D}};


\end{tikzpicture}
\caption{Optimal Contract} 

\label{fig:optimal_comp}
\end{center}
\end{figure}

\subsection{Solving the Revenue Maximizing Contract for a Piecewise Linear Cost Function}\label{sec:solvepiecewise}

The analysis in the previous section provides two useful insights into optimal contracts when the consumer has costly convex self-control preferences. First, we show that compromising contracts are strictly optimal when the cost function is strictly convex, and second, there is always loss of efficiency. However, convex cost function does not lend itself to comparative static analysis since there is no obvious analogue of the willpower parameter that controls the consumer's level of self-control. In this section, we propose a piece-wise linear and weakly convex cost function. The position of the kink is where the consumer's temptation cost starts increasing more rapidly and can be interpreted as the analogue of the willpower stock in the limited willpower model. Specifically, we solve for the revenue maximizing contract for the following specification.

\begin{equation}
 \varphi(x)=\begin{cases}
    lx & \text{if $x\leq w$}\\
    k(x-w) + lw & \text{if $x > w$}
  \end{cases}
\end{equation}
where $k>1>l > 0$.


\textsc{Indulging contract:}  In this case the monopolist offers $y^*$ at price $p(y^*)=u(y^*)$. We find  $p^{ind}$ by solving  (\ref{C_ind}) as
\begin{equation}
 p^{ind}=\begin{cases}
    \frac{u(x)+l(v(x)-e(y^*))}{1+l} & \text{if $e(x)-e(y^*)\leq (1+l)w$}\\
   \frac{u(x)+k(v(x)-e(y^*)-w)+lw}{1+k} & \text{if $e(x)-e(y^*) > (1+l)w$}
  \end{cases}
\end{equation} 

We now show that the indulging contract is better than the commitment contract. Suppose $x \neq y^*$  so that the indulging contract is distinct from the commitment contract. Assume  $e(x)-e(y^*)\leq (1+l)w$. By definition, 
\[
v(x)-u(x) > e(y^*) 
\]
if and only if
\[
u(x)+l(v(x)-e(y^*))>(1+l)u(x).
\]  if and only if indulging contract is strictly better.

\[
v(x)-u(x) >e(y^*) .
\]

Assume   $e(x)-e(y^*)> (1+l)w$, which implies  $e(x)-e(y^*) - w > 0$ if and only if 
$k(v(x)-u(x)-e(y^*) - w )> 0$ if and only if $u(x) + k(v(x)-e(y^*) - w ) +lw > ku(x) +u(x)$
if and only if
\[
u(x)+k(v(x)-e(y^*)-w)+lw>(1+k)u(x)\] 
if and only if the indulging contract is strictly better.   Hence the indulging contract is strictly better than the commitment contract.

\begin{figure}[h!]  
\begin{center}
\begin{tikzpicture}[scale=1.2] 
\draw [<->] (0,5) -- (0,0) -- (5,0);
\node [right] at (5,0) {\scriptsize{$v$}};
\node [above] at (0,5) {\scriptsize{$p$}};
\draw [blue, dotted] (0,.5) to (.5,0); 
\draw [blue, dotted] (0,2.5) to (2.5,0);
\draw [blue, dotted] (0,2.86) to (2.86,0);
\draw [blue, dotted] (0,3.7) to (3.7,0);
\draw [blue, dotted] (0,4.5) to (4.5,0);
\node [below] at (4.5,0) {\scriptsize{$e(z^*)$}};
\node [below] at (.5,0) {\scriptsize{$e(y^*)$}};
\node [below] at (2.4,0) {\scriptsize{$e(x^u)$}};
\node [below] at (3.7,0) {\scriptsize{$e(x^v)$}};
\node [below] at (2.9,0) {\scriptsize{$e(x)$}};


\draw[<-, Cyan] (.5,0) -- (2,.5) -- (4,2.5) ;
\draw[<-, blue, thick] (.5,0) -- (1.5,1) -- (3,1.5) ;
\draw[Cyan] (1.76,1.1) -- (3.26,1.6)--(5.26,3.6);



\draw[red] (1.44,1.06) to [out=0,in=-90] (2.13,1.57);
\draw[red] (2.13,1.57) to [out=95,in=310] (.9,3.8);
\draw[red] (1.44,1.06) to [out=180,in=305] (0.1,1.6);

\draw[fill] (0.5,0) circle [radius=0.05];
\draw[fill] (3,1.5) circle [radius=0.05];
\draw[fill=red, red] (1.76,1.1) circle [radius=0.05];
\node [above] at (1.76,1.1) {\scriptsize{C}};


\end{tikzpicture}
\caption{Optimal Contract} 

\label{fig:optimal_comp}
\end{center}
\end{figure}
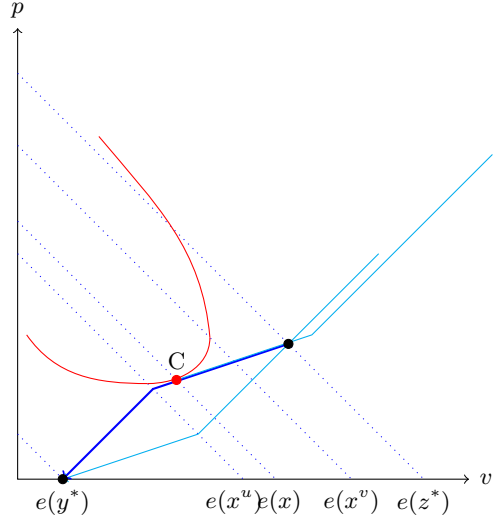

\textsc{Compromising Contract:}  To solve for the compromising contract, we first need to solve for $p(z^*)$ from (\ref{eq:priceofz}), which gives us: \begin{equation}
 p(z^*)=\begin{cases}
    \frac{u(z^*)+l(v(z^*)-e(y^*))}{1+l} & \text{if $e(z^*)-e(y^*)\leq (1+l)w$}\\
   \frac{u(z^*)+k(v(z^*)-e(y^*)-w)+lw}{1+k} & \text{if $e(z^*)-e(y^*) > (1+l)w$}
  \end{cases}
\end{equation} 

Substituting $p(z^*)$ in each range into (\ref{eq:revmaxprice}) and  solving the resulting equation, we get
\begin{equation}
 p^{comp}=\begin{cases}
    \frac{u(x)+lv(x)}{1+l} + \frac{k-l}{(1+k)(1+l)}e(z^*) - \frac{k}{1+k}e(y^*) -\frac{k-l}{1+k}w  & \text{if $e(z^*)-e(x)\leq (1+l)w$}\\
   \frac{u(x)+kv(x)}{1+k} - \frac{k}{1+k}e(y^*) & \text{if $e(z^*)-e(x) > (1+l)w$}
  \end{cases}
\end{equation} 

We next compare the compromising and the indulging contracts. If $e(z^*)-e(y^*) \leq (1+l)w$ (which implies $e(x)-e(y^*) \leq (1+l)w$),
the indulging and the compromising contracts generate the same revenue. This happens since the most tempting alternative cannot ``block" the ``bait alternative" which is $y^*$. Hence, w.l.o.g. the contract that maximizes revenue is an indulging contract and sells $x$ at price $$p^{ind}= \frac{u(x)+l(v(x)-e(y^*))}{1+l}. $$
%
%
%
%
If $e(z^*)-e(y^*) > (1+l)w$, comparing the revenue in all possible cases we see that the contract that maximizes revenue is a compromising contract and sells $x$ at price:
\begin{equation*}
 p^{comp}=\begin{cases}
    \frac{u(x)+lv(x)}{1+l} + \frac{k-l}{(1+k)(1+l)}e(z^*) - \frac{k}{1+k}e(y^*) -\frac{k-l}{1+k}w  & \text{if $e(z^*)-e(x)\leq (1+l)w$}\\
   \frac{u(x)+kv(x)}{1+k} - \frac{k}{1+k}e(y^*)  & \text{if $e(z^*)-e(x) > (1+l)w$}
  \end{cases}
\end{equation*} 

\subsubsection{Optimal Contract}
Next we find which alternative the monopolist should
sell to maximize its profit. From the set $\{ x :e(z^*) - e(x) \leq (1+l)w\}$, it is optimal to sell the maximizer of $\frac{u(x)+lv(x)}{1+l} - c(x)$. From the set $\{ x :e(z^*) - e(x) \geq (1+l)w\}$, it is optimal to sell the maximizer of $\frac{u(x)+kv(x)}{1+k} - c(x)$. Hence the optimal contract sells either 
$$\argmax_{x:e(z^*) - e(x) \leq (1+l)w}\frac{u(x)+lv(x)}{1+l}-c(x)\text{ \ or \ }  \argmax_{x:e(z^*) - e(x) \geq (1+l)w}\frac{u(x)+kv(x)}{1+k}-c(x)
$$
whichever generates the higher profit. Hence we get the following result.

\begin{proposition}\label{prop:mainapp_convex}
\begin{enumerate}
\item For any $e(z^*) - e(x_k) \geq (1+l)w$, the optimal contract is the best compromising contract selling $x_k$ at $$p(x_k)=\frac{u(x_k)+kv(x_k)-ke(y^*)}{1+k}$$. 

\item For any $e(z^*)-e(x_k) < (1+l)w < e(z^*)-e(x_l)$, the optimal contract is the best compromising contract, which sells an alternative between $x_l$ and $x_k$. 

\item For any $e(z^*)-e(x_l) \leq (1+l)w < e(z^*)-e(y^*)$, the optimal contract is the
best compromising contract that 
sells $x_l$ at $$p(x_l)=\frac{u(x_l)+lv(x_l)}{1+l} + \frac{k-l}{(1+k)(1+l)}e(z^*) - \frac{k}{1+k}e(y^*) -\frac{k-l}{1+k}w.$$

\item For any $e(z^*)-e(y^*) \leq (1+l)w$, the optimal contract is the indulging contract
selling $x_l$ at $$p(x_l)=\frac{u(x_l)+l(v(x_l)-e(y^*))}{1+l}.$$ 
\end{enumerate}
\end{proposition}

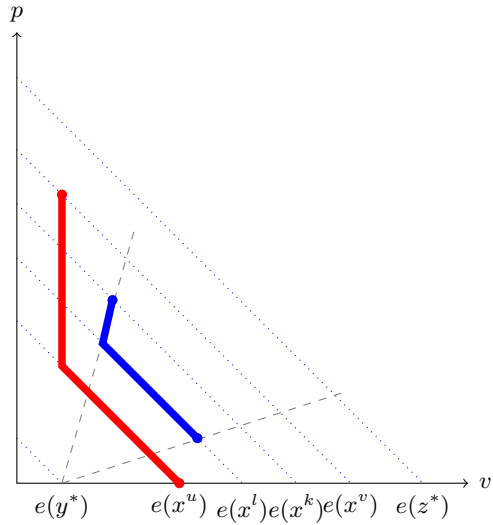
\begin{figure} [h!] 
\begin{center}
\begin{tikzpicture}[scale=1.2] 
\draw [<->] (0,5) -- (0,0) -- (5,0);
\node [right] at (5,0) {\scriptsize{$v$}};
\node [above] at (0,5) {\scriptsize{$p$}};
\draw [blue, dotted] (0,.5) to (.5,0); 
\draw [blue, dotted] (0,1.8) to (1.8,0);
\draw [blue, dotted] (0,2.5) to (2.5,0);
\draw [blue, dotted] (0,3.1) to (3.1,0);
\draw [blue, dotted] (0,3.7) to (3.7,0);
\draw [blue, dotted] (0,4.5) to (4.5,0);
\node [below] at (4.5,0) {\scriptsize{$e(z^*)$}};
\node [below] at (.5,0) {\scriptsize{$e(y^*)$}};
\node [below] at (1.8,0) {\scriptsize{$e(x^u)$}};
\node [below] at (3.7,0) {\scriptsize{$e(x^v)$}};
\node [below] at (3.1,0) {\scriptsize{$e(x^k)$}};
\node [below] at (2.5,0) {\scriptsize{$e(x^l)$}};



\draw[dashed, gray] (.5,0) -- (1.3,2.8);
\draw[dashed, gray] (.5,0) -- (3.6,1);

\draw[fill=red, blue] (1.06,2.03) circle [radius=0.05];
\draw[fill=red, blue] (2,0.5) circle [radius=0.05];
\draw[line width=1mm, blue] (1.06,2.03) -- (.95,1.55) -- (2,0.5);

\draw[fill=red, red] (.5,3.2) circle [radius=0.05];
\draw[fill=red, red] (1.8,0) circle [radius=0.05];
\draw[line width=1mm, red] (.5,3.2) -- (.5,1.3) -- (1.8,0);


%
%
%
%


\end{tikzpicture}

\caption{Optimal Contract for different levels of $w$}\label{fig:contractcurve}

\end{center}
\end{figure}

Now, we shall consider how the optimal contract, the profit, and the naive
consumer's welfare changes as the consumer's willpower changes. In Figure \ref{fig:contractcurve}, for comparison we provide the contract curve for the limited willpower model that we derived in \cite{MNOzdenoren19}, given by the red line. The blue line is the contract curve for the piece-wise linear model.  For any point on the contract curve, by moving down the iso-$e$ line, we can find out the product sold to the consumer by the optimal contract. The $y$-coordinate of the point gives us the price of this product in excess of its utility value. Hence as we move down the contract curve monopolist's profit declines and consumer's welfare increases.  

When $w$ is below a certain level, the product sold under the optimal contract and its price remain the same, which indicates that a small increase in $w$ does not help the consumer at all. This happens on the upper-left corner of the contract curve in Figure \ref{fig:contractcurve} which corresponds to $w \in [0,\frac{e(z^*)-e(x^k)}{1+l}]$. In this range, since monopolist sells $x^k$ at the same price,  both the monopolist's profit and consumer's welfare do not change. 

As the $w$ increases we enter the range ($w \in (\frac{e(z^*)-e(x^k)}{1+l},\frac{e(z^*)-e(x^l)}{1+l})$) which corresponds to the positive-sloped portion of the contract curve. In this range the excess temptation of the product sold goes down from $e(x^k)$ to $e(x^l)$, its price and the monopolist's profit drop, and the consumer's welfare increases. 

The next range ($w \in (\frac{e(z^*)-e(x^l)}{1+l},\frac{e(z^*)-e(y^*)}{1+l})$) corresponds to the linear decreasing portion of the contract curve. Unlike the limited willpower model, here, exploitation requires inefficiency  since the optimal contract sells the inefficient alternative $x^l$ at an exploitative price exceeding its utility value. In this range the excess temptation of the product remains constant at $e(x^l)$, its price and the monopolist's profit drop, and the consumer's welfare increases. 

Lastly, at the lower-right hand corner of the contract curve ($w \in [\frac{e(z^*)-e(y^*)}{1+l}, \infty)$), the nature of the optimal contract changes since the monopolist sells the inefficient alternative $x^l$ using the indulging contract at an exploitative price. Independent of willpower,  the naivety always  hurts the consumer.

\bibliographystyle{plainnat}
\bibliography{choice}

\end{document}